\documentstyle[aaspptwo]{article}
\def\lsim{\lower.5ex\hbox{$\; \buildrel < \over \sim \;$}}
\def\gsim{\lower.5ex\hbox{$\; \buildrel > \over \sim \;$}}

\begin{document}

\title{Bending Instability of an Accretion Disc Around a Black Hole}

\author{D. Molteni$^1$, K. Acharya$^{2}$, Oleg Kuznetsov$^{3}$, D. Bisikalo$^{4}$, 
Sandip K. Chakrabarti$^{5,2}$} 

\affil{$^1$Department of Physics, University of Palermo, Palermo, Italy\\
$^2$ Centre for Space Physics, 114/v/1A Raja S.C. Mullick Rd., Kolkata 700047\\
$^3$ Keldysh Institute of Applied Mathematics, Moscow \\
$^4$ Institute of Astronomy, Moscow \\
$^{5,2}$ S. N. Bose National Centre for Basic Sciences, JD-Block, Salt Lake,
Kolkata, 700098, INDIA\\ 
e-mail: molteni@gifco.fisica.unipa.it, 
space\_phys@vsnl.com, kuznecov@spp.keldysh.ru, bisikalo@inasan.rssi.ru \& chakraba@boson.bose.res.in}

\begin{abstract}
We demonstrate that generically an accretion disk around a compact object
could have a new type of instability in that the accretion flow need not be symmetric
with respect to the equatorial plane even when matter is supplied 
symmetrically farther out. We find that this behaviour is mainly due to 
interaction of outgoing matter bounced off the centrifugal barrier and the 
incoming accretion. We believe that X-ray variability could be 
developed by this instability.
\end{abstract}


\noindent Accepted for publication in ApJ Letters

\section{Introduction}
Accretion onto a black hole is generally assumed
to be thin and axisymmetric (e.g.  Pringle, 1981).
In presence of a close companion, accretion disk can develop
non-axisymmetric instability and spiral shocks are produced (Sawada,
Matsuda \& Hachisu, 1986). Bending waves and warps are known 
to be produced in planetary rings (Greenberg \& Brahic, 1990) 
and galactic disks (e.g., Pringle, 1997).
These instabilities are due to self-gravity of the disk or due to 
gravity of a nearby companion. In the present {\it Letter}, 
we demonstrate that an accretion disk could develop
a new kind of instability, because of interaction 
between accreting  matter and winds and outflows generated from the star or the
accretion disk.  The interface between the two becomes unstable 
due to shear instability or Kelvin-Helmholtz
instability. In presence of rotation, which is invariably true in the case of an accretion
disk, the instability is not the same in both the upper-half and the lower-half. As a result, the 
accretion process may not be symmetric with respect to the equatorial plane. 
The instability is present independent of polytropic index or cooling 
mechanism in the flow. 

In the past, a large number of numerical simulations which concern 
accretion flows around black holes, have been presented in the literature
(e.g., Hawley, Wilson \& Smarr 1984; Eggum, Coroniti \& Katz, 1985; Molteni,
Lanzafame \& Chakrabarti, 1994; Molteni, Sponholz \& Chakrabarti, 1996; hereafter MSC96
Ryu, Molteni \& Chakrabarti, 1997; Igumenshchev, Abramowicz, Narayan, 2000). 
These work assumed a priori that the flow has a 
equatorial symmetry and therefore the flow behaviour  was studied 
on one half  and used the `reflection' boundary condition on the
equatorial plane. With the advent of faster computational techniques
and computers, one could perform numerical simulations in more realistic cases,
namely, in both the halves simultaneously without using such an artificial 
boundary condition. MSC96 studied oscillating shocks to explain 
Quasi-Periodic Oscillations, but these oscillations 
are generated only in a part of the parameter space.
The vertical oscillation of the disk  we report here
occurs in a much larger range of the disk parameters,
only amplitude and frequency vary. For very large energy or very low angular momentum
when the infall velocity is high, but outflow is weak, this phenomenon
disappears. Thus, these new results may also cause X-ray variabilities 
in galactic black holes sources and optical/UV variabilities in Active Galaxies and
Quasars. A preliminary report on various shock instabilities are presented in
Molteni et al. (2001).

In the next Section, we present the formulation of the problem in both Smoothed
Particle Hydrodynamics (SPH) and in TVD method.
In \S 3, we present the numerical simulation results and demonstrate the 
instabilities. In \S 4, we discuss our results in light of shear instability. Finally,
in  \S 5, we draw conclusions. 

\section {Formulation of the Problem}

It has been discussed quite extensively in the recent literature  that matter
close to a black hole must be sub-Keplerian (e.g.,Chakrabarti, 1996).
This is because the boundary condition on the horizon forces the flow to radially 
move with the velocity of light. Thus, matter moving in Keplerian disks farther out
must become sub-Keplerian as it approaches a black hole. (In wind accreting 
objects, accreting matter could be sub-Keplerian throughout.) This assumption is
in vogue in the literature in the context of other flow models 
(cf. Narayan \& Yi 1995, Narayan, Kato and Honma, 1997, Artemova et al. 2001;
See, however Armitage et al., 2001 and  Hawley, 2000, for recent treatment
of the physical conditions of canonical disks in the zone of the last stable orbit.)
In the simulations described below, we start with a sub-Keplerian flow. For convenience, 
we use $r_g=2GM_{BH}/c^2$ the Schwarzschild radius of the black hole, 
to be the unit of distance, $r_g/c$ to be the 
unit of time and $\rho_{ref}$, a reference density to be the unit of density of matter. 
Here, $G$ and $c$ are the universal gravitational constant and velocity of 
light respectively and $M_{BH}$ is the mass of the black hole. We use Paczy\'nski-Wiita
(1980) pseudo-potential to describe gravitational field around a Schwarzschild black hole.
In all our simulations, we use axisymmetry and flow both above and below
the equatorial plane.  The first assumption is justifiable,
especially because the black hole itself is strictly axisymmetric and inner boundary condition
forces the flow to be axisymmetric close to the hole. However, far away, our assumption
restricts us from studying non-axisymmetric nature of this instability. This can be
overcome only with three-D simulations we plan to perform in future.
We inject matter at the outer boundary symmetrically 
with respect to the equatorial plane. We supply the sound speed, radial velocity,
density and specific angular momentum at the outer boundary.
We use two numerical procedures, namely, Smoothed Particle 
Hydrodynamics (Monaghan 1985; MSC96)
and TVD method (a combined Lax-Friedrichs-Osher scheme suggested by Vyaznikov, Tishkin, \&
Favorsky, 1989). For a sub-Keplerian flow at a distance of $r=X$, the free-fall time scale is given 
by $t_{ff}=X^{3/2}$. Thus if the outer boundary where matter is injected is kept at
$X=50$, one obtains $t_{ff}\sim 350$. For a convincing result, the simulation must 
run at least ten times this time scale. 
We included bremsstrahlung cooling in the hydrodynamic code 
however we verified that the instability we discuss is not due to any cooling processes.
The code were tested with theoretical solutions. For instance,
in Molteni, Ryu \& Chakrabarti (1996) the 
codes were found to produce nonlinear features such as standing shocks
roughly when theoretical simulations predict them.

\section{Results }

All the simulations presented here are carried out with an ideal gas of polytropic index $\gamma=5/3$
and a black hole of mass $M_{BH}=10^8 M_\odot$.
Fig. 1 shows the results of the SPH method at four different times $t=3000,\ 5000,\ 7500,$ and $13000$
(time marked in each box). Each panel shows velocity vectors  in $X-Z$ (meridional) plane, $Z=0$
being the equatorial plane. To bring clarity, only one out of three arrows have been 
plotted (this is why there are `gaps', especially in the inflow area). The inflow parameters are:
radial velocity $v_r=-0.1240$, sound speed is $0.0249$, $\rho_{ref}=
1.25 \times 10^{-19}$ and $\lambda=1.95$ is the specific angular momentum. 
This $\rho_{ref}$ corresponds to an accretion rate of ${\dot M}=1.25\times 10^{-4} \ {\dot M}_{Edd}$.
We also include bremsstrahlung cooling process. This tends to cool matter whenever it
becomes denser and hotter. As a result, more matter `sticks' together and produce enhanced 
clumpiness. Other than this, the bending instability is produced even without cooling. 

There are several features which are to be noted. First, matter is found to be repeatedly 
`pinched' during accretion process at a regular distance of $2-3\ h(r)$, where $h(r)$ is the 
local height of the inflow. Second, matter bounces back from the centrifugal
barrier and generates an outflow (outward pointing arrows) part of which falls back on the 
disk and deposits energy and momentum on the disk.
Third, accretion flow bends very severely and often the amplitude 
of the bending wave exceeds ten Schwarzschild radius, at a distance of $10-20 r_g$. Fourth,
the flow switches upper-half to lower-half and vice-versa 
in a time scale of about $t_{sw}=10000$ in units of the
light crossing time of the black hole. For $M_{BH}=10^{8} M_\odot$,
this corresponds to a switching time of about four months! For a galactic black hole
of $M_{BH}=10 M_\odot$ this time scale would be exactly $1$ second.
While some of these features may be enhanced because of our finite location of the boundary
and also because we assume strictly axisymmetric system, we believe that general behavior
should be similar in a three-dimensional simulations as well.
Vortices produced asymmetrically are found to affect the outflow rates in upper and lower 
halves. As a result, the momentum flux in the jet itself may be quasi-periodically 
switched to high and low values. 

In Fig. 2, we use the TVD method. Here we inject matter at $r=500r_g$ and use
radial velocity $v_r=-0.0444$, sound speed is $2.83 \times 10^{-3}$, specific angular momentum
is $\lambda=1.90$ and reference density is $\rho_{ref}=10^{-19}$. 
We use $129 \times 129$ unequally spaced grids. Velocity field is
shown with arrows drawn at every alternate grids for clarity, except in 
regions $|x|<100$ and $|z|<100$ where fewer arrows are drawn.
Superposed on the velocity field are contours of constant density, the outermost one closest to 
the axis (or the uppermost/lowermost ones at the outer boundary)
being that of $\rho=0.25 \rho_{ref}$. Density  in each contour inward is 
increased by a factor of two. Disk is found to be bent at several regions 
by a different degree. 

We plot in Fig. 3 the total amount of matter
in the upper (solid) and the lower (dashed) halves as a function 
of time (in units of $r_g/c$). 
Masses in the two halves vary in a complementary way. Large scale fluctuations are found
in all time scales, ranging from $1000$ to $100,000$ time units or more (certainly the
largest time-scale is related to the location of the boundary). In Fig. 4, we show
the variations of density (solid curves, in units of $\rho_{ref}$) 
and radial velocity (dashed curves, in units of velocity of light)
along the $Z-direction$ of the flow at three radial distanced 
$X=50,\ 150$ and $300$ (marked) respectively.  Radial velocity 
has been multiplied by a factor of $30$ to bring the plots in the same scale.
While quantities at $X=300$ are more or less symmetric with respect to
the equatorial plane ($Z=0$), quantities at $X=150$ have peaks in density and velocity
in the upper-half ($Z>0$) while those at $X=50$ have peaks in density and velocity 
in the lower-half ($Z<0$). The velocity jumps sharply from negative values (inflow) 
to positive values (outflow). Thus a very strong shear is operating on the surface 
of the disk. The effects of this are discussed below.

\section{Possible Origin of Bending of the Accretion Flow}

A {\it particle} of specific angular momentum $\lambda$ is bounced off a centrifugal
barrier located at $r$ where the centrifugal force balances gravity.
A flow has a pressure and therefore it bounces farther out. 
Outflowing wind and the incoming accretion interact and shear instability
could be developed. Chandrasekhar (1961, \S 105) 
provides a dispersion relation for a uniformly rotating flow of angular
velocity $\Omega$ (angular momentum vector pointing in the $Z$ direction)
streaming with velocities $v_1$ and $v_2$ along X-axis. For $\Omega=0$
this relation gives the perturbation frequency $\omega$ as a function of
wavenumber $k$ to be: $\omega= - k (v_1 +v_2)/2 \pm [-k_x^2 (v_1-v_2)^2/4]^{1/2}$
where we assumed density to be equal in both the media. The perturbation grows in time scale of,
$$
t_{KH} = \frac{2\pi}{\omega_i}=\frac{4\pi}{k (v_1-v_2)}=\frac{2 \lambda}{v_1-v_2} ,
\eqno{(1)}
$$
where, $\omega_i$ is the imaginary component of $\omega$ and $\lambda = 2\pi/k$
is the radial wave length.

From Figure 4, one can easily compute the growth time scale at various
locations. For instance, at $x=300,\ 150$ and $50$, for perturbation lengths comparable to
the local width of the disk, $\lambda_x \sim  80, \ 60$ and $40$, and velocity differences
$v_1-v_2 \sim 0.13,\ 0.17 $ and $0.25$ respectively, growth time scales
are $t_{KH} \sim 5000, \ 1850 $ and $350$ respectively. For a  black hole
of $M_{BH}=10^8 M_\odot$ as we are considering here, these time scales
correspond to  $14, \ 8.3$ and $3.5$ days respectively. The simulations were carried out
till $t=220,000$  and is much larger than the above time scales. 
For galactic black hole candidates ($M=10 M_\odot$), the times scales would be
$0.52$s, $0.185$s and $0.035$s respectively and oscillations of radiations
would have frequencies $\sim 2$Hz, $5.4$Hz and $ 28.5$Hz respectively. Thus
the effect of the oscillation is likely to produce a power spectra with 
excess power in around $2-30$Hz. We note that for low
$\lambda$, $v_1$ is large but the wind is absent.
For high $\lambda$, $v_1$ is smaller, but $v_2$ is large. So the frequency 
should be a function of the input parameters.
When rotation is present, $\omega$ is obtained not by analytical means, but by using 
the method of characteristics (Chandrasekhar, 1961).
When the axisymmetry is removed we still expect such instabilities, 
perhaps spiral in nature.

\section{Discussions and Conclusions}

We showed that in presence of outflows, there is a significant 
degree of time-varying distortion of the accretion disk surface. The interaction
of the oppositely moving flows suggest that the distortion or bending is 
mainly due to the shear instability, although local heating effects,
momentum deposition of the outgoing subsonic flow onto the disk, 
and local rotational effects may not be ruled out. Disk distortion is  found to be
a generic phenomenon provided both the disk and the jet form
and is present in all length scales whenever shear effects are important
whether cooling is present or not. We found that both Smoothed Particle Hydrodynamics 
and TVD Methods show this instability. The effects we discuss
are very similar to what Norman, Winkler \&  Smith (1982) and Chakrabarti (1988)
discussed in the context of instabilities in jets in presence of terminal Mach 
shock. There too back-flowing cocoon formed around a jet has a destabilizing
effect on the jet. In the case of jets, the back-flow is due to tangential discontinuity
separating the jet matter from that of the ambient matter. In the
present case, the back-flow is due to the centrifugal barrier.

Observationally we expect that such a behaviour is likely to produce time variabilities in
radiation emitted from the inner part (say, $r < 200-300r_g$) of the disk. Thus whereas
in AGNs,  optical and UV radiations could be seen in time scales of days, in galactic 
black hole candidates, soft and hard X-rays are likely to show variabilities 
in time scales of seconds or less.  Since the whole disk does not oscillate
coherently or quasi-coherently, the power density spectrum is likely to produce a
very broad bump due to this type of oscillations. 

{}

\vfil\eject
\begin{figure}
\caption {
Examples of the oscillation of the disk around the equatorial plane at different times
of SPH simulation. The meridional planes (X-Z) are shown. Distances are measured in units of
$r_g$ and the times are measured in units of $r_g/c$. The interaction between the
outflow and the inflow is clear in these pictures. One out of three arrows have been 
plotted for clarity. See text for parameters.}
\end{figure}

\begin{figure}
\caption {Contours of constant density superposed  on the velocity field for a case
where injection of matter takes place at $x=500$. Finite Element method has been used.
Alternate arrows are plotted for clarity except in regions $|x|<100$ and $|y|<100$
where number of arrows drawn are reduced further. Bending of the disk takes place
on all scales. }
\end{figure}

\begin{figure}
\caption{
Variation of mass of the matter in the upper and the lower-halves of the 
simulation for which Fig. 2 was drawn as a function of simulation time.
Variation is seen in time scales of $1000$ to as large as $100,000$
(in geometrical units).  }
\end{figure}

\begin{figure}
\caption{
Variation of density (solid) and velocity (dashed)
at various distance (marked on each curve)
from the black hole as a function of the vertical distance
$Z$. Note that while the velocity and density are
roughly symmetrical around $Z=0$ for $X=300$, 
they are asymmetrically distributed
and $X$ decreases. For $X=150$ peaks in density
and velocity lie in the
upper-half, and for $X=50$ the peaks lie in the
lower-half.
}
\end{figure}
 
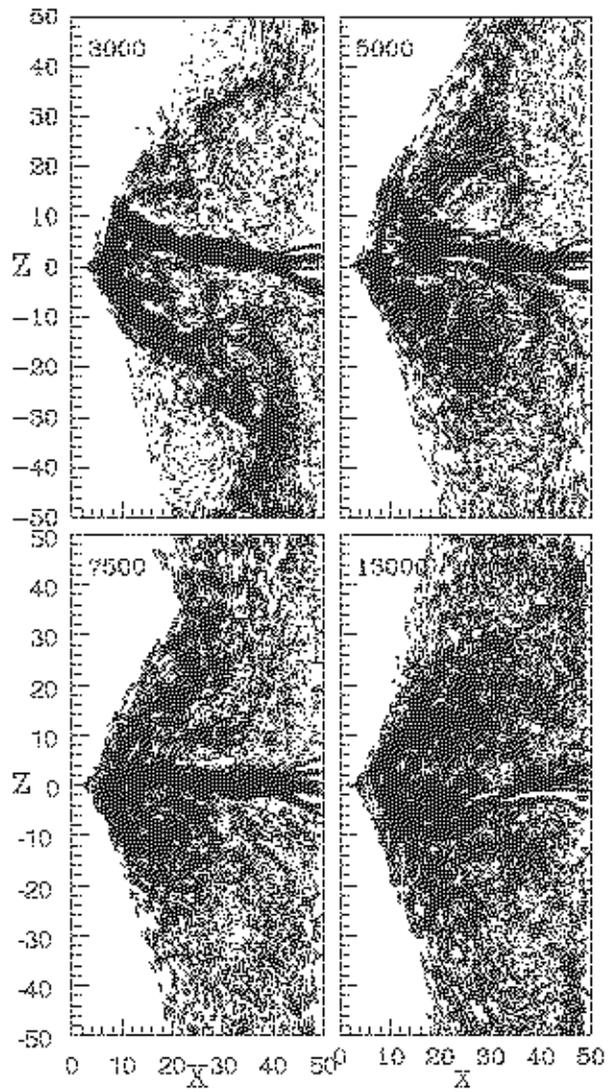
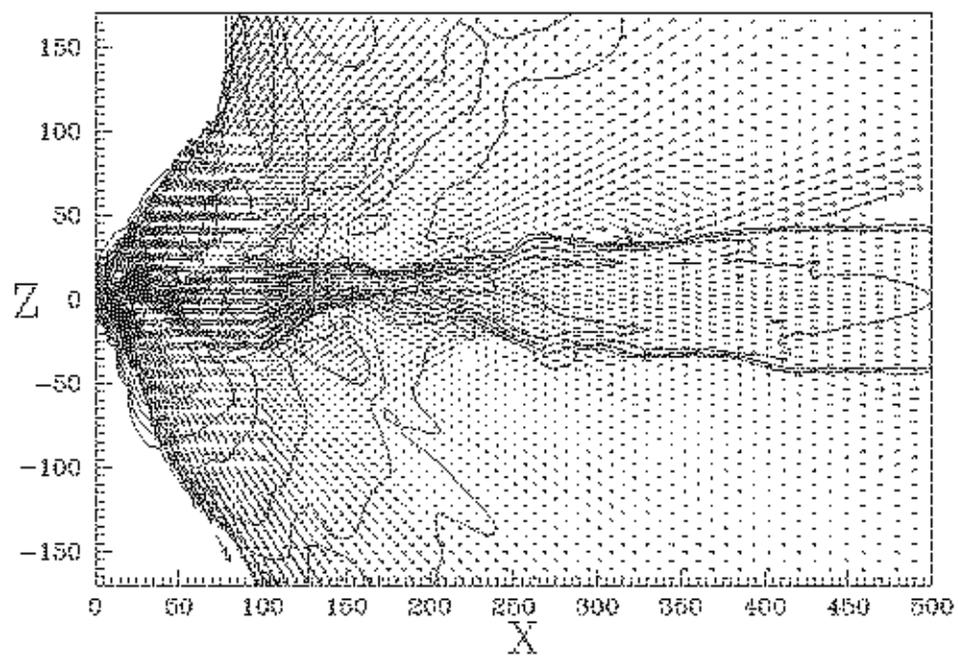
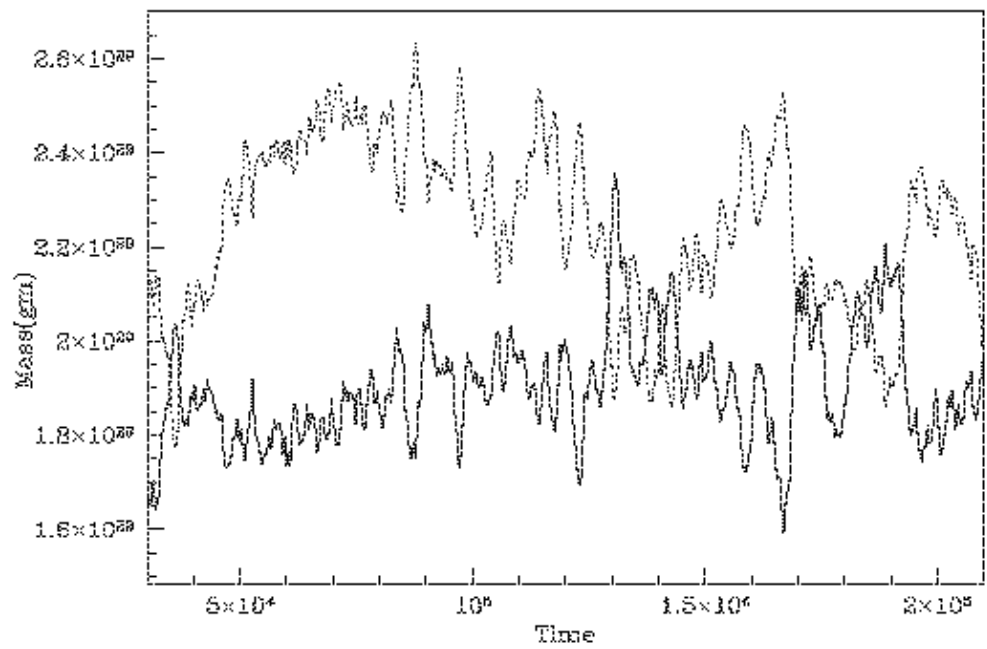
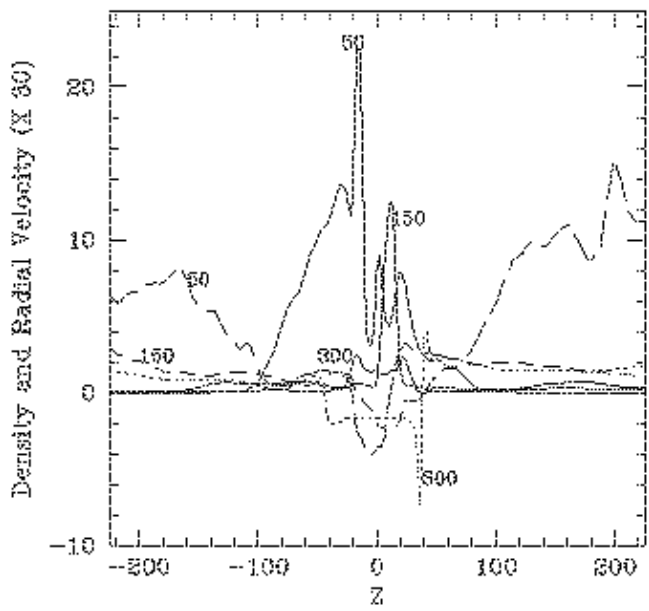
\end{document}